# Optical Combs and Optical Vortices Combined for Spatiotemporal Manipulation of Light and Matter

Akifumi Asahara,[1,2] Satoru Shoji,[1,2] and Kaoru Minoshima[1,2,a]

[1]*Department of Engineering Science, Graduate School of Informatics and Engineering, The University of Electro-Communications (UEC), 1-5-1 Chofugaoka, Chofu, Tokyo 182-8585, Japan*

[2]*JST, ERATO MINOSHIMA Intelligent Optical Synthesizer (IOS) Project, 1-5-1 Chofugaoka, Chofu, Tokyo 182-8585, Japan*

a) Author to whom correspondence should be addressed. Electronic mail: k.minoshima@uec.ac.jp

**ABSTRACT**

In this study, we demonstrate the concept of combining optical combs with optical vortices for the first time. By combining the advantages of the both light sources, we realize an "optical vortex comb" technology for arbitrary spatiotemporal phase manipulation. This idea corresponds to simultaneous control of the longitudinal and transverse modes of light based on the high controllability of the optical comb. As a proof-of-concept experiment, we applied an optical vortex comb to the generation and rotational control of a ring-shaped optical lattice. Furthermore, to demonstrate the utility of this technique, a rotational optical manipulation of microspheres was demonstrated using the optical ring lattice as an optical tweezer light. Here, we present a new applicability of the optical comb by utilizing its characteristics far beyond the conventional range.

## 1. Introduction

Optical frequency combs have realized a wide range of innovative applications in metrology as high-quality next-generation light sources with extremely high precision, coherence, and controllability [1–3]. The optical comb is widely known as a "precision frequency ruler". Its features have especially been used in the precision measurement field where high absolute frequency accuracy is required, such as in frequency standards, time transfer, and optical lattice clocks. However, recently its applications have expanded broadly beyond the common conventional sense. For instance, the dual-comb spectroscopy technique is a successful application of the optical comb, which realizes unprecedented high-speed and precise spectroscopic measurements using two asynchronous optical combs [4–7]. In the early stages of the study, gas spectroscopy and distance measurements have been main research targets. However, in recent years, a wide variety of ideas using dual-comb techniques have been developed, such as solid-state characterization [8], ultrafast time-resolved measurements [9,10], coherent two-dimensional spectroscopy [11], two-photon spectroscopy [12], nonlinear Raman spectroscopy [13], imaging measurements [14], coherent modulation techniques [15], and others. In addition, remarkable progress has been made in the development of optical comb light sources, such as micro-combs [16,17] and single cavity dual-comb sources [18–20]. The optical comb field is expected to develop further and grow as a key technology in metrology.

Herein, we focus on the combination of the optical vortex as a novel direction in the optical comb studies. The optical vortex is light with a characteristic helical spatial phase distribution, also known as a Laguerre-Gaussian beam [21,22]. The optical vortex has many features, such as the donut beam profile and orbital angular momentum (OAM) transfer via photons, which have attracted much attention in the research community. A variety of interesting studies using the unique characteristics of optical vortices have been reported, such as spatial division multiplexing communications, super-resolution imaging, and transverse mode quantum entanglement. A technology referred to as the "ring-shaped optical lattice" is also known among such optical vortex applications, which is a ring-shaped interference pattern generated by two superposed optical vortices with different helicities [23–27]. This technique is expected to be a novel light tool for optical tweezers that induce rotational optical manipulation.

In this study, we demonstrate the concept of combining an optical comb with an optical vortex for the first time. By combining the advantages of both these light sources, we realize an "optical vortex comb" technology for arbitrary spatiotemporal phase manipulation. This idea corresponds to the simultaneous control of the longitudinal and transverse modes of light based on the high controllability of the optical comb. As a proof-of-concept experiment, we applied an optical vortex comb to the generation and control of the ring-shaped optical lattice. Furthermore, to demonstrate the utility of this technique, a rotational optical manipulation of microspheres was demonstrated using the developed light. By utilizing the characteristics of the optical comb far beyond the conventional application range, a new applicability of the optical comb is presented.

## 2. Concept of "optical vortex comb"

Figure 1 shows the schematic of our proposed concept, "optical vortex comb," by combining the optical comb and optical vortex. The optical comb has two controllable frequency parameters, $f_{rep}$ and $f_{ceo}$, which denote the repetition rate and carrier envelope offset frequency, respectively. By stabilizing these frequencies, we can realize the optical comb as a precise frequency ruler and determine the absolute frequency of each longitudinal mode. From the perspective of the time domain, it is also known that the $f_{ceo}$ relates to the pulse-to-pulse variation of the carrier envelope phase, and the temporal phase is defined as $\phi_c = 2\pi f_{ceo} t$ [15]. This relationship ensures that a high $f_{ceo}$ controllability of the optical comb achieves ultraprecise and arbitrary manipulation of the temporal phase of light. On the other hand, the optical vortex has a transverse spatial phase distribution expressed as $\phi_v = \ell\theta$, where $\ell$ denotes the topological charges of a Laguerre–Gaussian beam, and $\theta$ is the azimuthal angle in the plane orthogonal to the propagating axis. The $\ell$ parameter determines the helicity of the optical vortex, corresponding to $2\pi\ell$-phase rotation along the azimuth.

We focus on the high affinity of these two light sources. Since these two lights have phase controllability in the orthogonal dimensions along the temporal and spatial axes, these states are simultaneously compatible. Therefore, by combining the concepts, the spatiotemporal phase of light can be controlled by the precise optical comb controllability via the spatial phase structure of the optical vortex. This can be defined as a novel light concept, "optical vortex comb". This idea extends the application of optical combs towards new dimensions far beyond the conventional research.

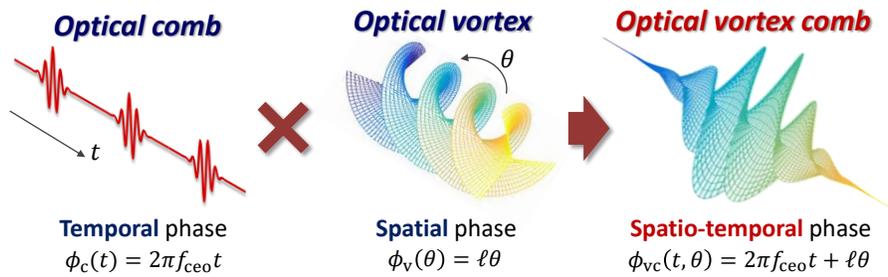

Fig. 1. Concept of the "optical vortex comb" by combining optical combs with optical vortices. The spatiotemporal phase of light is controlled by the optical comb parameter, $f_{ceo}$. The phase offsets are omitted for the simplicity of the model.

## 3. Ring-shaped optical lattice generation and control using optical vortex combs

### 3.1 Principle

To demonstrate the utility of the proposed concept, we perform generation and control of the ring-shaped optical lattice. This is a point-symmetric interference pattern generated by the superposition of two optical vortices with different topological charges, $\ell_1$ and $\ell_2$. The technique has attracted attention as a tool for rotational optical manipulations [23] and an experimental platform of one-dimensional physics with a periodic boundary condition [24,25]. Although continuous-wave lasers are commonly used, new techniques have been realized using ultrashort pulse lasers, such as optical tweezers using nonlinear effects [26] and femtosecond ultrafast rotation of the ring lattice [27]. Here by introducing the precise frequency controllability of optical comb, arbitrary control of the rotational motion is possible.

As shown in Fig. 2, the symmetry of the generated ring lattice is determined by the difference of the topological charges, $\Delta\ell = |\ell_1 - \ell_2|$, and the spatial profile is rotated by the phase difference between the two optical vortex combs, $\Delta\phi_{vc} = \phi_{v1} - \phi_{v2}$. Hence, when a combination of two coherent optical vortex combs, i.e., a "dual-optical vortex comb," is applied as the light source, the rotational movement of the ring lattice is controlled through an extended phase relationship,

$$\Delta\phi_{vc}(t, \theta) = 2\pi\Delta f_{ceo}t + \Delta\ell\theta. \qquad (1)$$

By considering the symmetry of the spatial structure, the angular velocity of the optical ring lattice is determined as $\omega = 2\pi\Delta f_{ceo}/\Delta\ell$. Notably, the spatiotemporal phase is freely controlled based on the precise frequency controlof the $\Delta f_{ceo}$. This indicates that the high temporal phase controllability of the optical comb is directly converted into the spatial dimension via the structure of the optical vortex.

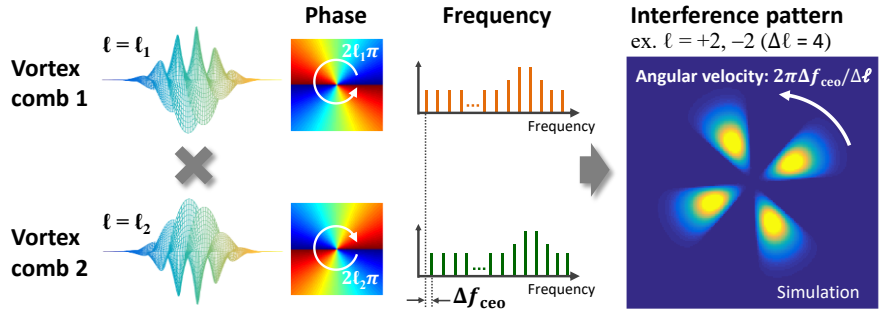

Fig. 2. Generation and control of the ring-shaped optical lattice using two optical vortex combs with different topological charges and $\Delta f_{ceo}$. The simulated optical ring lattice when $\Delta\ell = 4$ is described in the right panel.

### 3.2 Experimental system

We developed an experimental system presented in Fig. 3(a). A laboratory-built amplified Er-fiber comb system was used as an optical comb source, where $f_{rep}$ was 56.5 MHz, and the center wavelength was 1,560 nm (192 THz). The amplified power was several tens of mW, and the duration was broadened by a few picoseconds due to the chirp in the transmission fibers. The frequencies of $f_{ceo}$ and $f_{rep}$ were phase-stabilized to a microwave frequency standard; see [8,15] for details of our comb system. The output was incident to an acousto-optic modulator (AOM) (Crystal Technology, Model 3165-1) employed as a frequency shifter. The incident comb was split into transmitted and diffracted combs. As shown in Fig. 3(b), the frequency spectrum of the diffracted comb was shifted according to $f_{AOM}$ induced on the AOM. Therefore, these two combs can be treated as an effective dual-comb light source with an adjustable $\Delta f_{ceo}$ and identical $f_{rep}$.

To generate two optical vortex combs with different topological charges, we used a q-plate as shown in Fig. 3(c). The q-plate is a well-known optical converting plate and has the functionality of generating an optical vortex from the incident circular polarized light [28], where a right circular Gaussian beam ($\ell = 0$, $\sigma = +1$) is converted into a left circular vortex beam ($\ell = +2$, $\sigma = -1$) and vice versa. This phenomenon is interpreted as the conversion between the spin angular momentum and OAM of light. In the present system, two optical vortex combs were generated from two inverted circular polarization combs. In Fig. 3(a), the output of the combs from the AOM were superposed on the polarizing beamsplitter (PBS) after adjusting their polarizations into orthogonal states. The powers of the combs were adjusted to be the same using a variable neutral density filter, and the comb pulses were temporally overlapped using a piezo delay line. The overlapped dual-comb was transmitted through a single-mode fiber for matching their transverse modes and securing the interference visibility. The polarizations of the outputs were carefully adjusted into opposite right and left circular states ($\sigma_1 = +1$ and $\sigma_2 = -1$) using waveplates. Finally, two optical vortex combs with opposite topological charges ($\ell_1 = +2$ and $\ell_2 = -2$, so $\Delta \ell = 4$) and opposite polarizations ($\sigma_1 = -1$ and $\sigma_2 = +1$) were generated using a q-plate (Thorlabs, WPV10-1550). For interference of the dual-optical vortex combs, a PBS was inserted to take the projection of their polarizations. Behind the PBS, one path was used to observe the ring-shaped optical lattice using an InGaAs near-infrared camera (HAMATSU Photonics, C12741-03), and another path was used for an optical manipulation demonstration which is mentioned later.

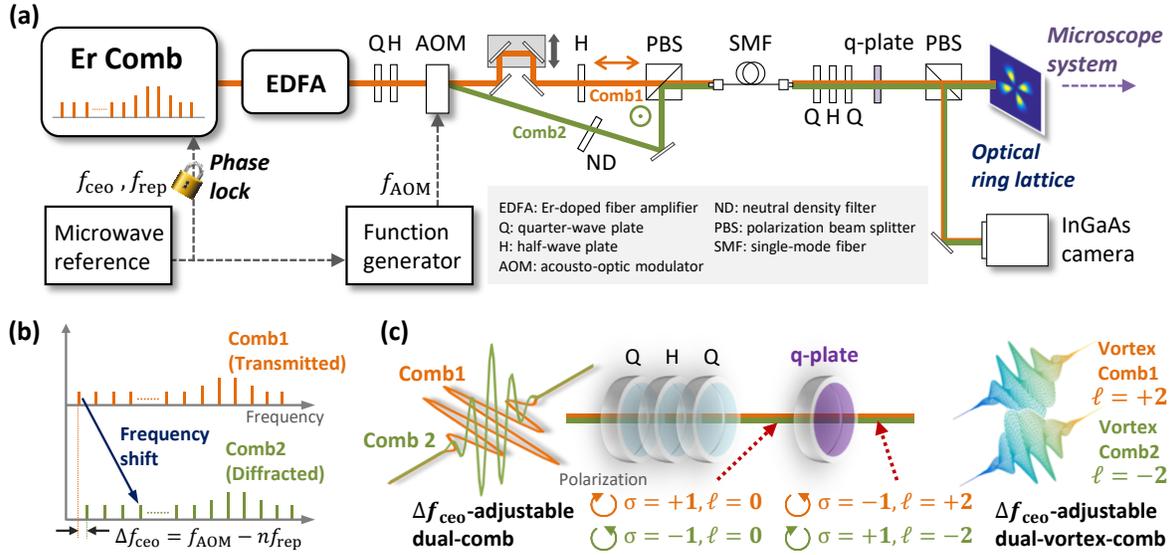

Fig. 3. (a) Experimental system for generation and control of a ring-shaped optical lattice using optical vortex combs. (b) Generation of an effective dual-comb light with an adjustable $\Delta f_{ceo}$ using AOM. (c) Principle of dual-vortex comb conversion using a q-plate with different topological charges and an adjustable $\Delta f_{ceo}$.

### 3.3 Demonstration of the ring-shaped optical lattice control

Figure 4 shows the experimental results. A series of observation images of the generated ring-shaped lattice are shown in Fig. 4(a). Here, a four-fold symmetric optical ring lattice was clearly observed. This spatial structure reflects the topological charge difference, $\Delta\ell = 4$, of the generated optical vortex combs, and the observed images are in good agreement with the simulated result. In the experiment, the $\Delta f_{ceo}$ of the two vortex-combs was set accurately at 1 Hz using the AOM and function generator. As a result, we succeeded in manipulating the rotational motion of the optical ring-shaped lattice via the frequency parameter of the optical combs.

In Fig. 4(b), as a two-dimensional color plot of the interference intensity along the time and azimuth, we summarized the spatiotemporal change of the optical ring lattice generated when $\Delta f_{ceo}$ was 1 Hz. The analysis was performed for the data at the radius where the four interference peaks were located. The slope in this color plot corresponds to the angular velocity of the rotational movement of $2\pi$ [rad/s], which is determined by the theoretical relationship, $\omega = 2\pi\Delta f_{ceo}/\Delta\ell$. The cross section on the left panel shows the intensity distribution along the azimuth direction at $t = 1$ s, reflecting the clear four-fold symmetrical spatial structure. On the other hand, the cross section on the top panel shows the temporal evolution of the interference intensity at the azimuth of $\pi/4$, reflecting the accurate and periodic rotation of the optical ring lattice. The visibility of these interference patterns was 0.9. In the developed system, using the perfect coaxial alignment with the SMF, we achieved a lattice pattern with good symmetry and high interference visibility.

By taking advantage of the high controllability of the optical combs, it is possible to arbitrarily control the rotational movement. The $\Delta f_{ceo}$ parameter can be easily changed via the settings of the function generator driving the AOM, and the angular velocity and rotational direction can be manipulated. For example, as shown in the upper part of Fig. 4(c), a rapid and invert rotation could be realized by setting the $\Delta f_{ceo}$ as –2 Hz. In principle, angular velocity control can be realized in a wide frequency range from microhertz to megahertz. Furthermore, when frequency modulation is induced to $\Delta f_{ceo}$, it is also possible to realize more complicated rotational movements, such as swing motion and stop-and-go motion. For example, the lower part of Fig. 4(c) shows the result of $\Delta f_{ceo}$ modulation using a sine function with a ±4-Hz amplitude change and 0.5-Hz modulation frequency. The wave-shaped color plot indicates that a swing motion was successfully induced in the optical ring lattice.

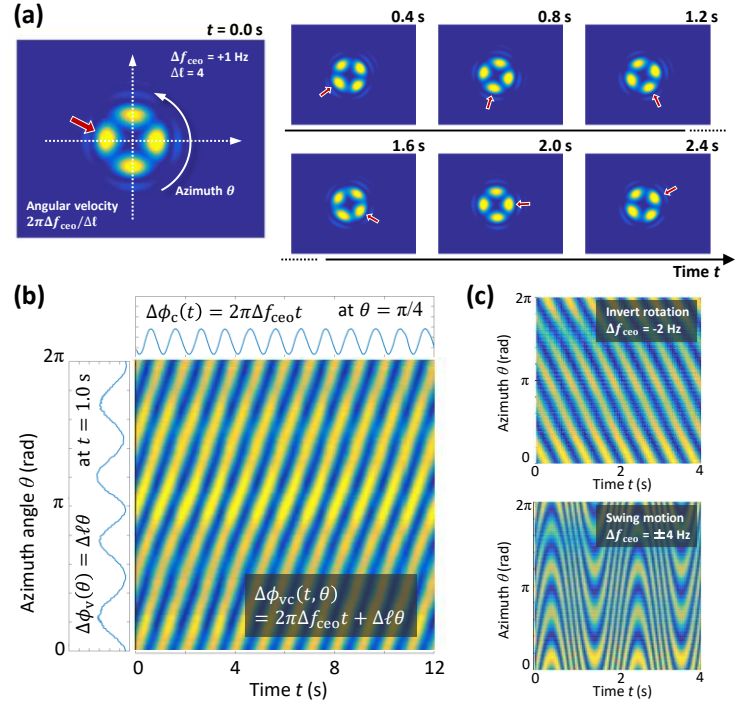

Fig. 4. (a) Series of observed images of the optical ring lattice generated when $\Delta f_{ceo}$ was 1 Hz. The red arrows indicate an identical rotating petal. (b) Spatiotemporal change of the interference intensity along the time and azimuth directions. (c) Spatiotemporal changes in the rapid and invert rotation ($\Delta f_{ceo} = -2$ Hz) and sine modulated swing motion ($\Delta f_{ceo} = \pm 4$ Hz, 0.5-Hz modulation frequency).

## 4. Rotational manipulation of matter using optical vortex combs

Furthermore, to demonstrate the applicability of the developed optical vortex comb concept, we applied rotating light to the optical manipulation. As shown in Fig. 5(a), we constructed a dry-mount microscopic system for the optical tweezers experiments. The generated four-fold symmetric optical ring lattice ($\Delta\ell = 4$) was tightly focused using an objective lens with an NA of 0.85. The optical power at the focused spot was 2.4 mW. The measurement target was polystyrene beads with a diameter of 2 μm, which is commonly used in optical tweezers. The beads were dispersed in water and sandwiched between two glass plates with a space of 150 μm. The microscopic images were observed using a CCD camera with white illumination light applied from the bottom.

Figure 5(b) shows the results of the optical manipulation experiment. Since a tightly focused optical field works as a tweezing light, in the case of an optical ring lattice, there are trapping spots corresponding to the number of $\Delta\ell$. In this experiment, we succeeded in the simultaneous optical trapping of four polystyrene beads using the four-fold symmetric optical ring lattice. In addition, by applying $\Delta f_{ceo}$ of 0.8 Hz, the rotational manipulation of the trapped microbeads was successfully demonstrated, which is clearly seen from the series of images in Fig. 5(b). As described above, using $\Delta f_{ceo}$ modulation, it is possible to create a more complicated operation.

These experiments demonstrated that the high temporal phase controllability of the optical combs could be effectively converted to control the spatial phase via optical vortices. To the best of our knowledge, these studies are the first applications of optical combs to the rotational control of light and matter. Conventionally, such spatial light operation has been achieved using mechanical movements such as a galvanometer mirror. By introducing an optical comb, it is expected that stable and high-speed all-optical manipulation can be realized without mechanically moving parts. Moreover, because the optical vortex comb is a type of ultrashort pulse laser, there is a possibility that the developed light source is applied to the nonlinear optical tweezer [27] utilizing its high peak power. In addition, we have dealt with only the limited case of $\Delta\ell = 4$; however, it is also possible to arbitrarily select the spatial structure by changing the topological charge distributions of the generated vortices. By actively controlling the optical comb frequency parameters in the time domain as well as the optical vortex parameters in the spatial domain, we believe that a wider variety of versatile optical manipulation systems can be designed.

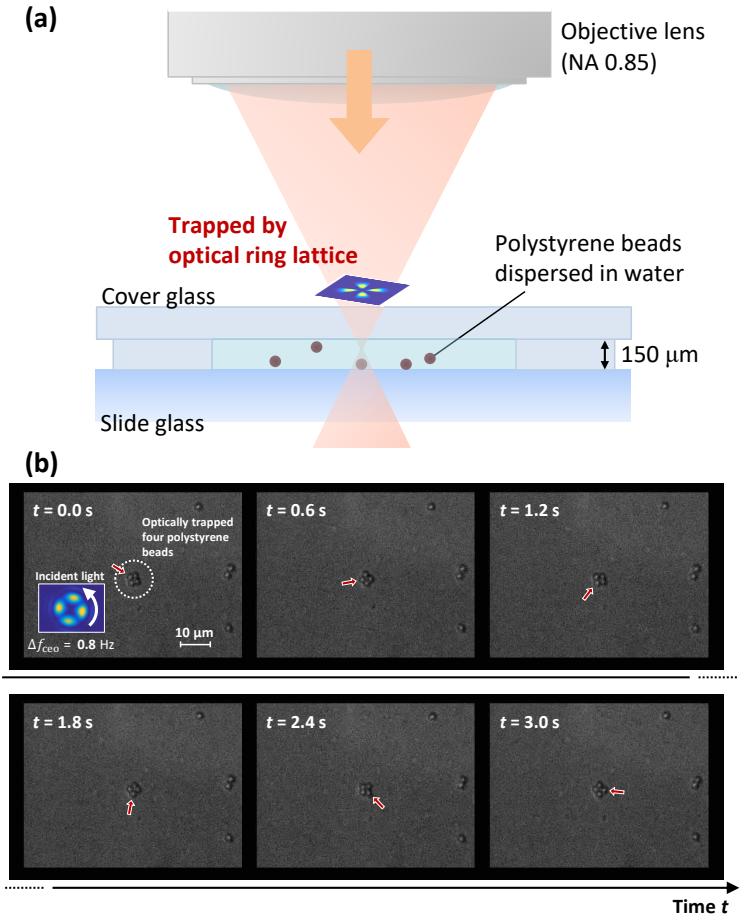

Fig. 5. (a) Microscopic system constructed for the optical manipulation experiment using the optical ring lattice. (b) Series of observed images of the rotating microbeads, that were trapped with the optical ring lattice based on the dual-optical vortex comb with $\Delta f_{ceo}$ of 0.8 Hz. The red arrows indicate an identical rotating microsphere.

## 5. Conclusion

We proposed a novel concept, an optical vortex comb, that enables spatiotemporal phase control by combining optical combs with optical vortices in this study. As demonstrative experiments, the rotational movement (angular velocity, rotational direction, and modulation) of the four-fold symmetric optical pattern was successfully controlled via the $\Delta f_{ceo}$ change. Furthermore, the arbitrary rotational manipulation of the microspheres was demonstrated using rotating optical tweezing light. Through these successful experiments, the applicability of the proposed concept was demonstrated. This proof-of-principle study will lead to advanced multi-dimensional applications of optical combs and optical vortices in which the spatiotemporal coherence of light is actively utilized.

**Acknowledgments**

This work was supported by JST, ERATO MINOSHIMA Intelligent Optical Synthesizer Project (JPMJER1304) and partly supported by JSPS Grant-in-Aid for Young Scientists (B) (JP17K14322).